\documentclass[letterpaper, 10 pt, conference]{ieeeconf}  

\IEEEoverridecommandlockouts                              

\usepackage{graphics} 
\usepackage{epsfig} 
\usepackage{mathptmx} 
\usepackage{amsmath} 
\usepackage{amssymb}  

\usepackage[normalsize,it,hang]{subfigure}
\newtheorem{remark}{Remark}  
\usepackage{algorithm}
\usepackage{algpseudocode}

\usepackage{xcolor}

\title{\LARGE \bf
Detection and suppression of epileptiform seizures \\ via model-free control and derivatives in a noisy environment}

\author{C\'edric Join\textsuperscript{1,6}, D. Blair Jovellar\textsuperscript{2,3}, Emmanuel Delaleau\textsuperscript{4} and Michel Fliess\textsuperscript{5,6}
\thanks{$^{1}$CRAN (CNRS, UMR 7039)), Universit\'{e} de Lorraine, BP 239, 54506 Vand{\oe}uvre-l\`{e}s-Nancy, France. 
\emph{@: cedric.join@univ-lorraine.fr}}%
\thanks{$^{2}$Department of Neurology \& Stroke, University of T\"ubingen, T\"ubingen, Germany.
\emph{@: desiree-blair.jovellar@uni-tuebingen.de}}
\thanks{$^{3}$Hertie-Institute for Clinical Brain Research, University of T\"ubingen, T\"ubingen, Germany.}
\thanks{$^{4}$ ENI Brest, IRDL (CNRS, UMR 6027), 29200 Brest, France. \emph{@: delaleau@enib.fr}}
\thanks{$^{5}$  LIX (CNRS, UMR 7161), \'Ecole polytechnique, 91128 Palaiseau, France. 
\emph{@: Michel.Fliess@polytechnique.edu, michel.fliess@swissknife.tech}  }
\thanks{$^{6}$ AL.I.E.N., 7 rue Maurice Barr\`{e}s, 54330 V\'{e}zelise, France. 
\emph{@: \{cedric.join, michel.fliess\}@alien-sas.com}}
}

\begin{document}

\maketitle
\thispagestyle{empty}
\pagestyle{empty}

\begin{abstract}

Recent advances in control theory yield closed-loop neurostimulations for suppressing epileptiform seizures. These advances are illustrated by computer experiments which are easy to implement and to tune. The feedback synthesis is provided by an intelligent proportional-derivative (iPD) regulator associated to model-free control. This approach has already been successfully exploited in many concrete situations in engineering, since no precise computational modeling is needed. iPDs permit tracking a large variety of signals including high-amplitude epileptic activity. Those unpredictable pathological brain oscillations should be detected in order to avoid continuous stimulation, which might induce detrimental side effects. This is achieved by introducing a data mining method based on the maxima of the recorded signals. The  real-time derivative estimation in a particularly noisy epileptiform environment is made possible due to a newly developed algebraic differentiator. The virtual patient is the Wendling model, i.e., a set of ordinary differential equations adapted from the Jansen-Rit neural mass model in order to generate epileptiform activity via appropriate values of excitation- and inhibition-related parameters. Several simulations, which lead to a large variety of possible scenarios, are discussed. They show the robustness of our control synthesis with respect to different virtual patients and external disturbances.

\end{abstract}


\section{Introduction}


The complexity of brain explains why the \emph{closed-loop}, or \emph{feedback}, setting, which is a key concept in control theory, is so underdeveloped in computational neuroscience as analyzed in several recent publications (see, e.g., \cite{jungle,argentina,mar,c2b,tub}), in spite of some preliminary attempts (see, e.g., \cite{kassiri}). This is confirmed \cite{schu} by the secondary role of brain research in \emph{systems biology}, a discipline that focuses on the connections between biology and control systems (see, e.g., \cite{alon,murray}).

This communication reports some preliminary computer experiments on \emph{epilepsy}, an important neurological disorder, also known as a {\emph{seizure disorder}} \cite{fisher}: 
\begin{enumerate}
    \item A closed-loop seizure suppression is proposed. It is based on an \emph{intelligent} regulator derived from \emph{model-free} control (MFC) \cite{ijc,nicu}. This setting has already led to many concrete applications (see, e.g., references in \cite{ijc,nicu}). Let us emphasize here a most recent comparison \cite{its} with other popular types of control synthesis in some specific questions arising in intelligent transportation systems, where the superiority of MFC is asserted. See below some of its main features: 
    \begin{itemize}
        \item MFC does not necessitate any mathematical modeling, and therefore neither delicate parameter identification procedures.
        \item It is much easier to tune than proportional-integral-derivative (PID) controllers which are the most popular industrial feedback loops (see, e.g., \cite{astrom}).
        \item It has already been successfully illustrated in biomedicine and bioengineering \cite{bara,he,anesth,diab,ventil}.
    \end{itemize}
   Following \cite{nicu}, the use of an \emph{intelligent proportional-derivative} controller, or \emph{iPD}, seems more adapted to regulate, via electric stimulations, a large variety of high-amplitude epileptic activity.
    \item Those aberrant burst events recur unpredictability and often are separated by long interictal time lapses. A continuous stimulation like above might therefore induce detrimental side effects \cite{hungary}. A real-time seizure detection is thus necessary for triggering the above feedback. Frequency-domain techniques are today a crucial ingredient in most publications on this subject (see, e.g., \cite{detect1,detect2,detect3,yuan,zhou}, and references therein). We follow here another quite recent route using algebraic manipulations in the time-domain (see \cite{easy,mexico,linidentif}, 
 for the theoretical background and the presentation of successful applications), and more specifically, here, via a new data mining viewpoint~\cite{agadir} where the maxima of the recorded signal are determined. Let us summarize this approach:  
    \begin{itemize}
        \item Maxima in a seizure are large and close to each other.
        \item Using derivatives for seizure detection, which involves a noisy environment, is a known challenge in engineering. 
        \item Our \emph{algebraic differentiator}, which is borrowed from \cite{mboup} (see, also, \cite{othmane}), does not necessitate any probabilistic and/or statistical assumption on the noise corruption.
    \end{itemize}
    \item The \emph{virtual} patient, i.e., the epileptiform signal, is provided by Wendling's neural mass model \cite{wendling1}. This computational model, which is now recognized as an efficient approach to get insights into this disease, is deduced from the well-known Jansen-Rit set of ordinary differential equations \cite{jansen} (see also \cite{epilept,wendling2}). The stimulators location  mimics \cite{arrais20,arrais19}. Modifying the many parameters of Wendling's model makes it easy to verify the inherent robustness of our model-free control. 
\end{enumerate}

Our paper is organized as follows. After a short presentation in Sect. \ref{mod} of Wendling's model where some stimulators are added, Sect. \ref{seiz} summarizes the derivatives estimation in a noisy environment and proposes a seizure detection algorithm, which is illustrated by computer experiments. After a review of some basic aspects of model-free control, Sect. \ref{clos} is discussing several scenarios on seizure suppression. See Sect. \ref{conclu} for short concluding remarks. 



\section{Model}\label{mod} 
\begin{figure*}[!ht]
\centering%
{\epsfig{figure=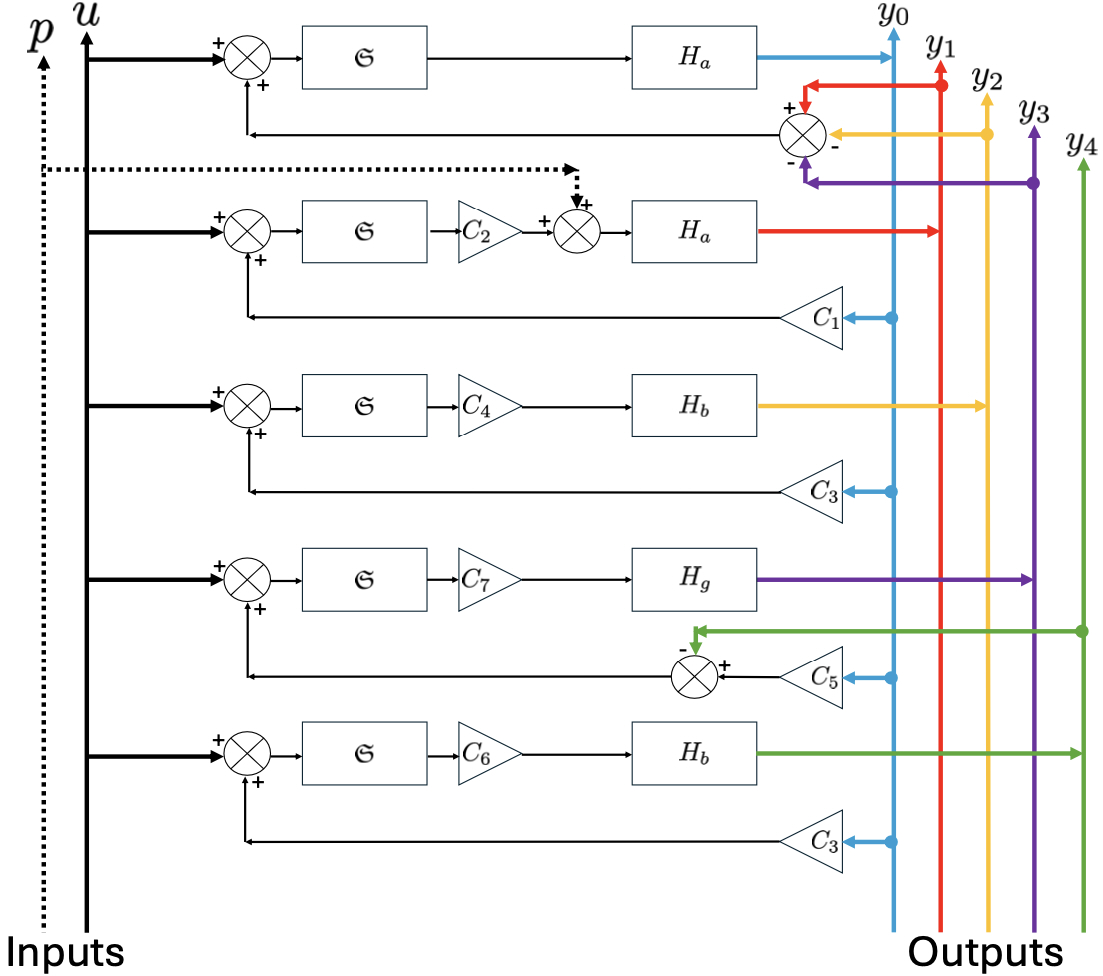,width=0.59\textwidth}}
\caption{Model scheme: block diagrams}\label{sc}
\end{figure*}

The following modification of Wendling's model \cite{wendling1}, which is due to \cite{arrais20,arrais19}, is obtained by adding stimulation (see \cite{arrais21} for a slightly different choice):  \begin{equation}\label{model}
\begin{cases}
\ddot y_0=Aa\mathfrak{S}(u+y_1-y_2-y_3)-2a\dot y_0 -a^2 y_0\\
\ddot y_1=Aa\left(p+C_2\mathfrak{S}(u+C_1y_0)\right)-2a\dot y_1-a^2y_1\\
\ddot y_2=BbC_4\mathfrak{S}(u+C_3y_0)-2b\dot y_2-b^2 y_2\\
\ddot y_3=GgC_7\mathfrak{S}(u+C_5y_0-y_4)-2g\dot y_3-g^2y_3\\
\ddot y_4=BbC_6\mathfrak{S}(u+C_3y_0)-2b\dot y_4-b^2y_4

\end{cases}
\end{equation}
where
\begin{itemize}
    \item $A, a, B, b, G, g,C_1, \dots, C_7$ are positive constants; 

     \item     \textit{A}, \textit{B} and \textit{G} represent the amplitude of average excitatory (EPSP), slow and fast inhibitory (IPSP) postsynaptic potentials, respectively.
    %
    \item  {The variables $y_0$, $y_1$, $y_2$, $y_3$, $y_4$ represent the post-synaptic potentials of pyramidal cells, excitatory feedback, and both slow and fast inhibitory interneurons, respectively. }   
    \item $p(t)$ is an external perturbation corresponding to excitatory inputs from neighboring areas. There are various representations in the existing literature:  white Gaussian noise, constant value, sum of them\dots
    \item The electrical stimulation $u(t)$ is the control variable. This follows the assumption that the stimulation generates an electric field with a direct de- or hyperpolizing linear effect onto the mean membrane potential of the neuronal subsets \cite{radman}.
    \item The sigmoid function $\mathfrak{S}(\square)$ reads 
    $$
    \begin{aligned}
    \mathfrak{S}(\square)=& \frac{v_{\rm max}}{2} \left(1 + \tanh{\frac{r}{2}(\square - v_0)} \right)\\ =& \frac{v_{\rm max}}{1+\exp(r(v_0-\square))}
    \end{aligned}
    $$
    where $v_{\rm max}$ is the maximum firing rate, $v_0 $ is the average membrane potential acting as a firing threshold, $v_{\rm max}$ is the maximum firing rate, and $r > 0$ is a constant.
   \end{itemize}
See Fig. \ref{sc} for a block diagram description  of~\eqref{model}, where  $H_a=\frac{Aa}{s^2+2as+a^2}$, $H_b=\frac{Bb}{s^2+2bs+b^2}$, $H_g=\frac{Gg}{s^2+2gs+g^2}$ are transfer functions. The summation $y_m=y_1-y_2-y_3$ represents the incoming firing rate to the population of pyramidal cells.  

\section{Seizure detection}\label{seiz}
\subsection{Differentiation in a noisy environment}

Here we will present a simplified version, borrowed from~\cite{easy}. Consider the polynomial function $\mathfrak{p}_1(t) = a_0 + a_1 t$, $t\geqslant 0$, $a_0, a_1 \in \mathbb{R}$. Classic \emph{operational calculus} (see, e.g., \cite{yosida}), or \emph{Laplace transform}, yields 
$P_1 = \frac{a_0}{s} + \frac{a_1}{s^2}$.
Multiply both sides by $s^2$: 
\begin{equation}\label{s}
s^2 P_1 = a_0s + a_1
\end{equation}
Deriving both sides with respect to $s$, which corresponds in the time domain to the multiplication by $- t$, yields $a_0$:
\begin{equation}\label{a0}
  a_0 = s^2 \frac{dP_1}{ds} + 2sP_1  
\end{equation}
Then $a_1$ is given by Eqn. \eqref{s}
\begin{equation}\label{a1}
  a_1 = -s^3 \frac{dP_1}{ds}-s^2P_1  
\end{equation}
In order to get rid of the positive powers of $s$, which corresponds in the time domain to derivatives w.r.t. time, multiply both sides of Eqn. \eqref{a0} (resp. \eqref{a1}) by $s^{-n}$, where the integer $n > 0$ is large enough. It yields in the time domain:
$$\frac{d\mathfrak{p}_1}{dt} = a_1=\frac{6}{T^3}\int_0^T\left(-T+2\tau\right)\mathfrak{p}_1(\tau)d\tau$$
where $T > 0$ is the window size used for the estimation. The integral, which is mitigating the corrupting noise in the sense of \cite{bruit}, may be in practice replaced by a digital filter.
The extension to polynomials of arbitrary degree is obvious and, therefore, also to truncated Taylor expansions. Details and references on the computer implementations may be found in \cite{othmane}, as well as references to many concrete applications.

\subsection{Seizure detection}

A seizure may be characterized by a short time lapse between two large maxima. Those maxima correspond to zero crossings of the derivative.

\subsubsection*{An academic illustration}
\begin{figure*}[!ht]
\centering%
\subfigure[\footnotesize $y_m$  ]
{\epsfig{figure=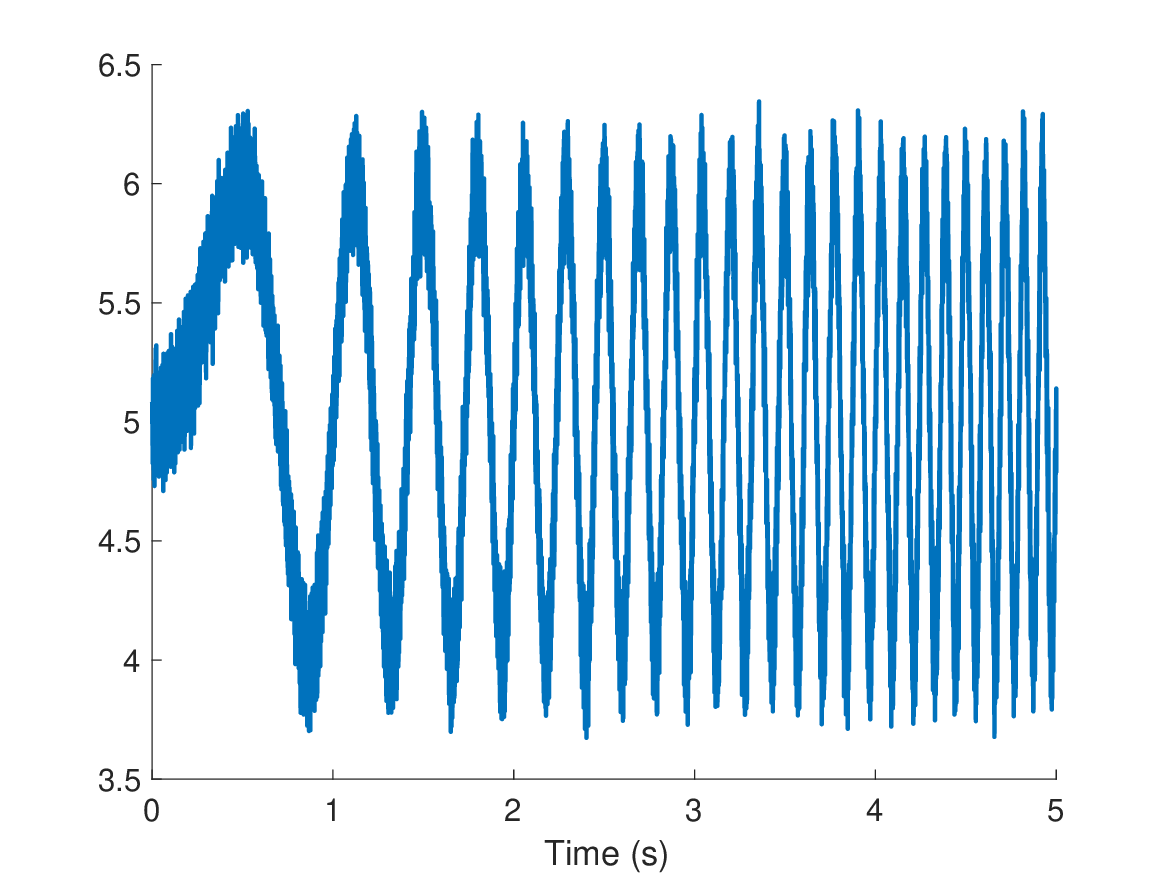,width=0.19\textwidth}}
\subfigure[\footnotesize Estimation of $\dot y_m$, zero crossing thresholds (- -) and (. -)  ]
{\epsfig{figure=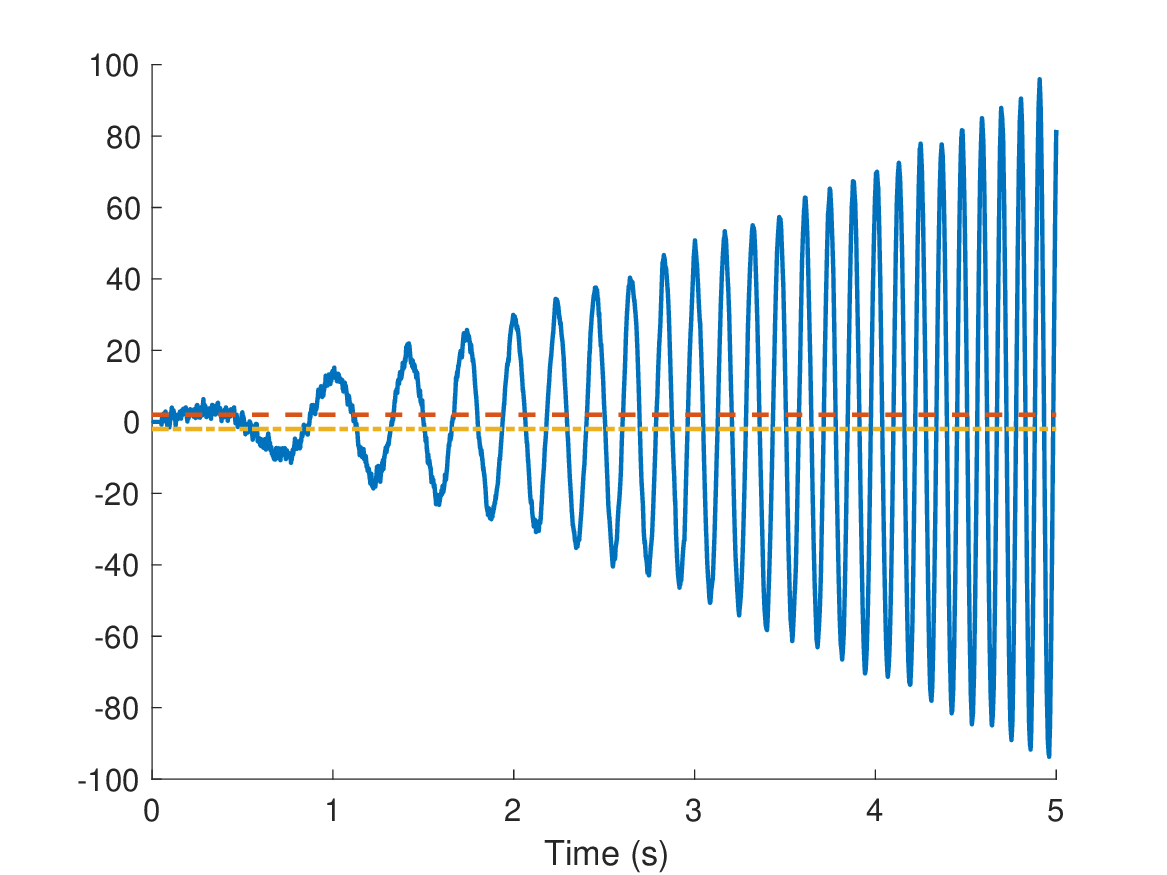,width=0.24\textwidth}}
\subfigure[\footnotesize Time between two maxima ]
{\epsfig{figure=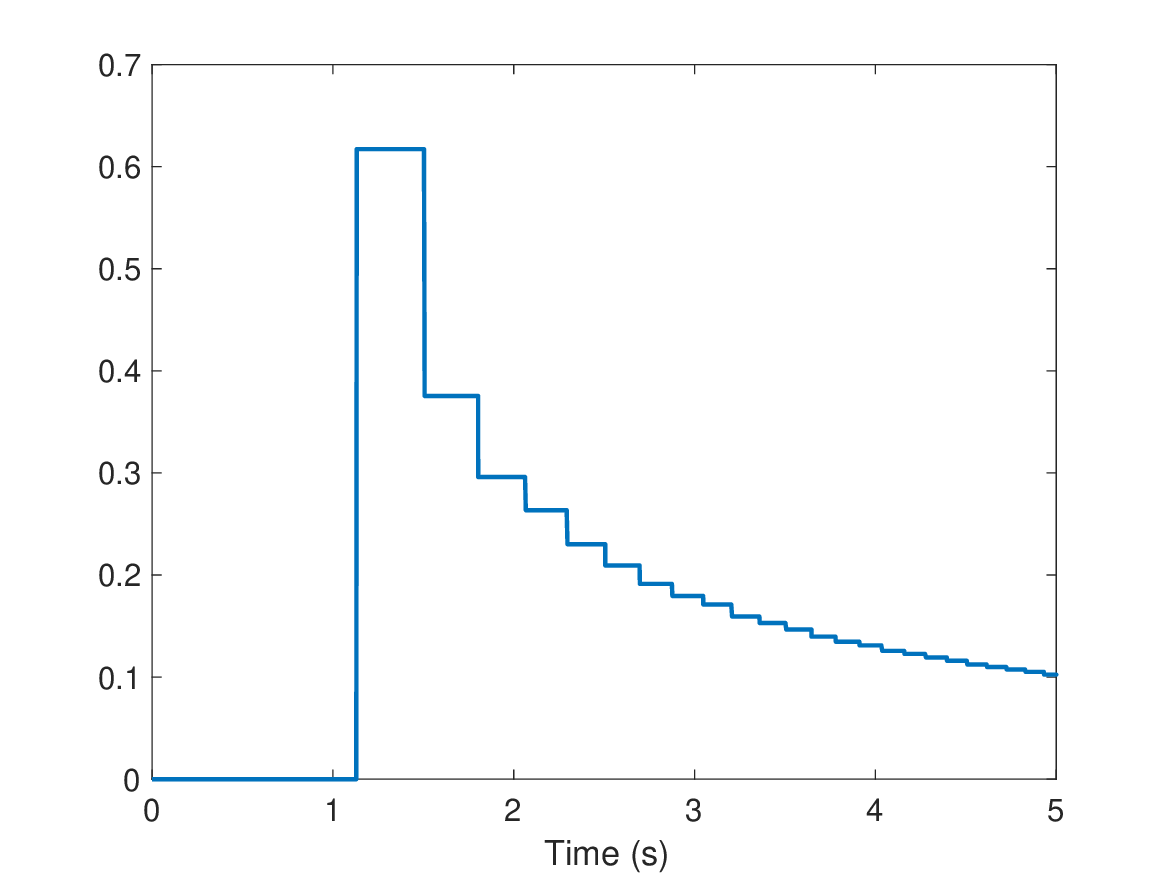,width=0.24\textwidth}}
\caption{Radar signal: Seizure detection}\label{Ci}
\end{figure*}

Take the \emph{chirp}, i.e., a frequency modulated signal, which is familiar in radar engineering (see, e.g., \cite{radar}),  
$$y(t)=\sin(2\pi t^2)+5, \quad \hspace{1cm} 0 \leqslant t \leqslant 5 $$
It is corrupted by an additive white Gaussian noise (mean: 0, standard deviation: 0.1).
Fig. \ref{Ci} exhibits excellent results.

\begin{figure*}[!ht]
\centering%
\subfigure[\footnotesize $y_m$  ]
{\epsfig{figure=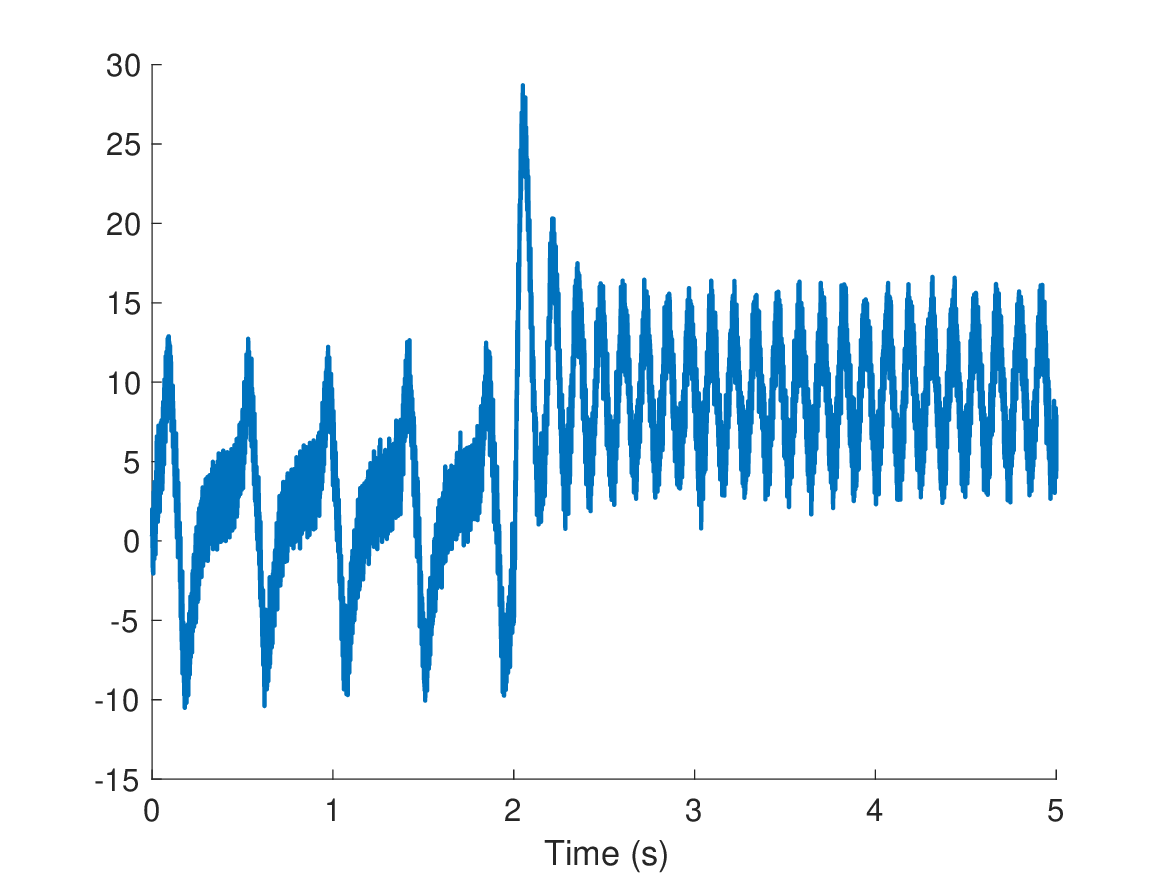,width=0.24\textwidth}}
\subfigure[\footnotesize Estimation of $\dot y_m$, zero crossing thresholds (- -) and (. -)  ]
{\epsfig{figure=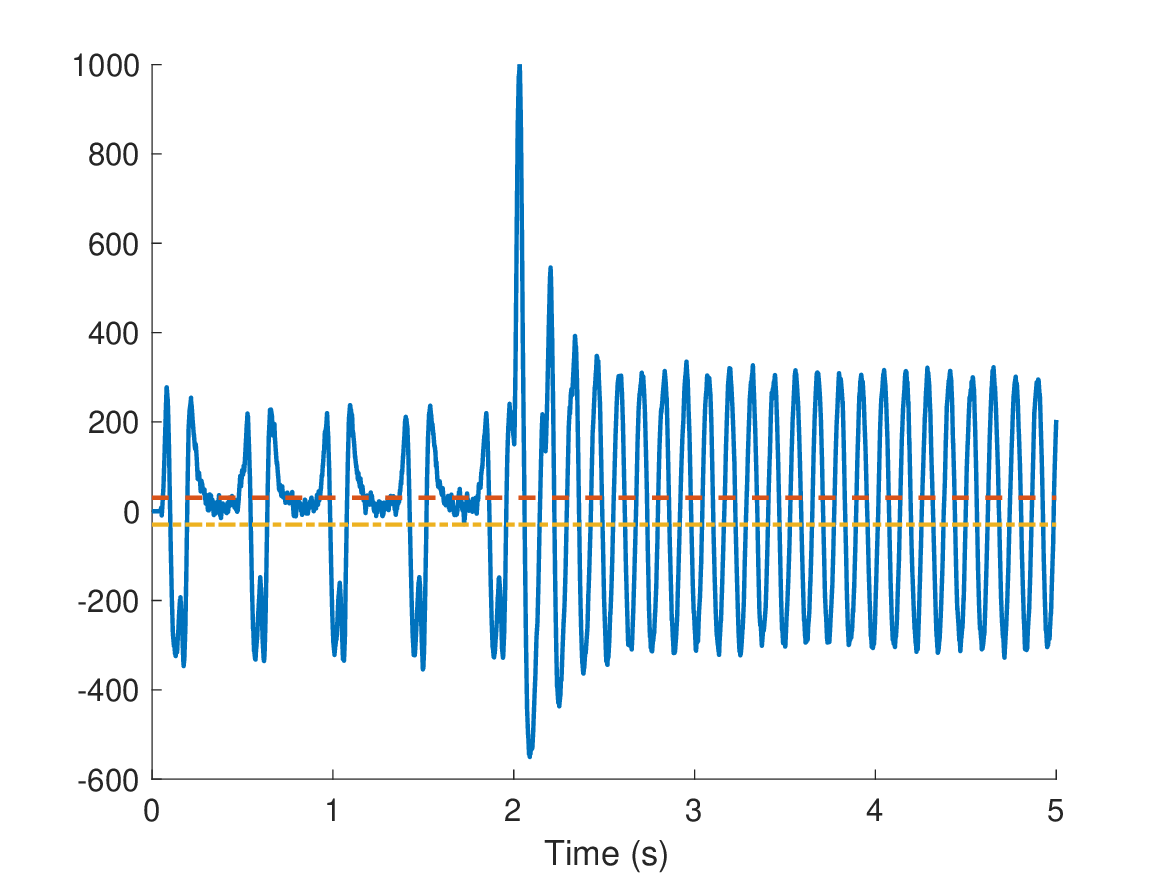,width=0.24\textwidth}}
\subfigure[\footnotesize Time between two maxima (--) and detection threshold (- -) ]
{\epsfig{figure=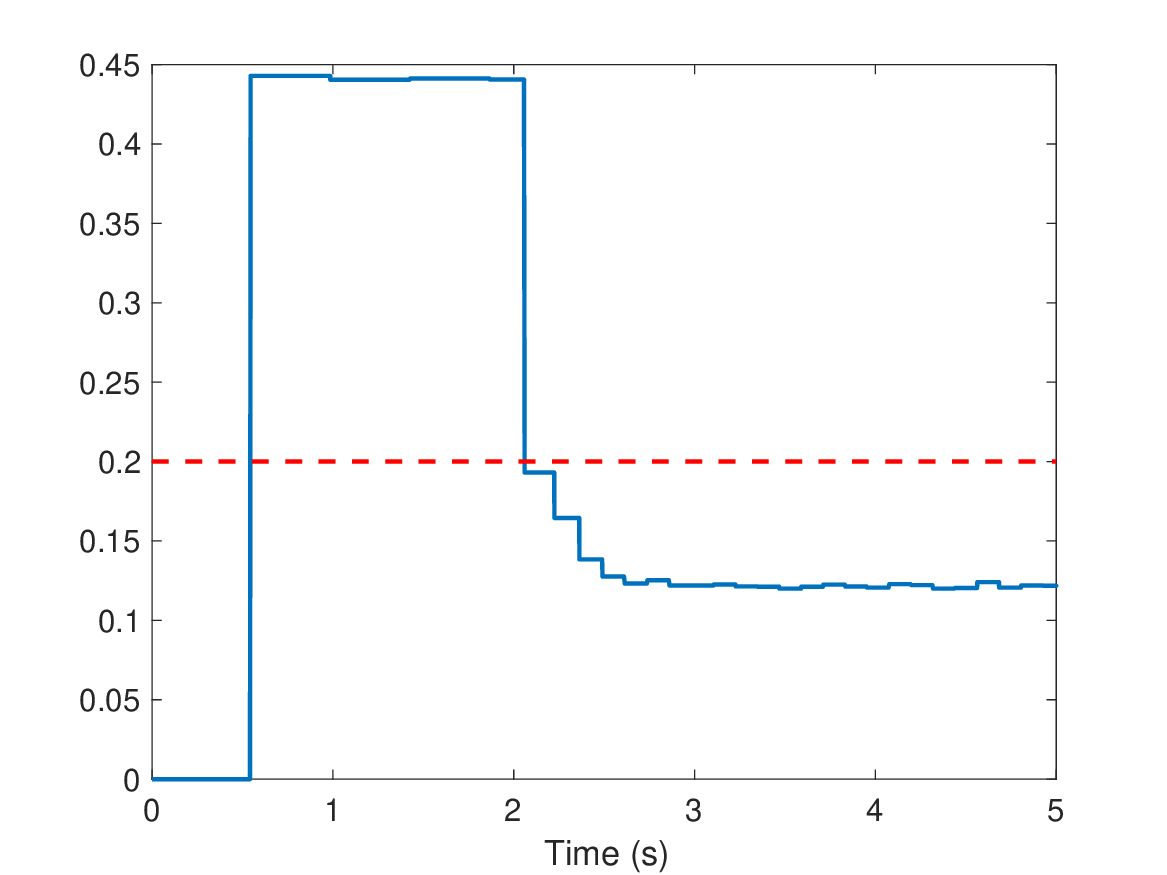,width=0.24\textwidth}}
\caption{Epilepsy: Seizure detection}\label{SD}
\end{figure*}

Fig. \ref{SD}-(a) displays a signal produced by Wendling's model~\cite{wendling1}, i.e., without stimulation, i.e., $u(t) = 0$, in Eqn.~\eqref{model}. Its right part shows a seizure, which ought to be detected in order to avoid any stimulation when there is no crisis \cite{hungary}. The derivative of the recorded signal $y_m (t)$ is reported in Fig.~\mbox{\ref{SD}-(b)}. See Fig. \ref{SD}-(c) for the time lapse between consecutive maxima. The choice of an adequate threshold \cite{agadir} for this time lapse yields the detection of the seizure.


\section{Closed-loop control}\label{clos}


\subsection{The ultra-local model}
Consider, following \cite{nicu}, the \emph{ultra-local model} of order $2$
\begin{equation}\label{ul2}
    \ddot{y}(t) = F(t) + \alpha u(t)
\end{equation}
\begin{itemize}
    \item $\alpha \in \mathbb{R}$, $\alpha \neq 0$, is a constant such that the three terms in Eqn. \eqref{ul2} are of the same magnitude. Therefore $\alpha$ does not need to be precisely determined.
    \item The \emph{data-driven} quantity $F$, which subsumes not only the poorly known system structure but also any external perturbation, may be estimated via the measurements of $u$ and $y$.
\end{itemize}

\noindent Introduce the \emph{intelligent proportional-derivative} controller, or \emph{iPD}, 
\begin{equation}\label{ipd}
    u(t) = -\frac{F_{\rm est}(t) - \ddot{y}^\star (t) + K_P e(t) + K_D \dot{e}(t)}{\alpha}
\end{equation}
where
\begin{itemize}
    \item $F_{\rm est}(t)$ is an estimate of $F$ (see Eqn.~\eqref{estim});
    \item $y^\star(t)$ is a \emph{reference} trajectory, and $e(t) = y(t) - y^\star(t)$ is the \emph{tracking} error;

\item $K_P, K_D \in \mathbb{R}$ are \emph{tuning} gains.
\end{itemize}   
Eqn. \eqref{ul2} and \eqref{ipd} yield
$
\ddot{e}(t) + K_D \dot{e}(t) + K_P e(t) = F(t) - F_{\rm est}(t)  
$.
Select $K_P$, $K_D$ such that the real parts of the roots of the polynomial $s^2 + K_D s + K_P$ are strictly negative. It ensures that $\lim_{t \to +\infty} e(t) \approx 0$ if the estimate $F_{\rm est}$ is ``good'' i.e., $F(t) - F_{\rm est}(t) \approx 0$. Thus local stability around the reference trajectory is trivially ensured via this feedback loop.
The following estimate of $F$ in Eqn. \eqref{ul2} is borrowed from~\cite{nicu}:
\begin{equation}\label{estim}
\begin{aligned}
  F_{\rm est}(t) = &\frac{60}{\tau^5} \int_{0}^\tau(\tau^2 + 6\sigma^2 - 6\tau \sigma)y(t - \tau +\sigma) d\sigma \\ &- \frac{30 \alpha}{\tau^5} \int_{0}^\tau (\tau - \sigma)^2 \sigma^2 u(t - \tau +\sigma) d\sigma
  \end{aligned}
\end{equation}
\begin{itemize}
\item It is a real-time estimate: $\tau > {0}$ is ``small.''
\item It is indeed data-driven: the numerical value of $F_{\rm est}$ results from the knowledge of the control and output variables $u$ and $y$.
\item It is a ``low pass filter'' thanks to the integrals in Formula~\eqref{estim}: noise in the sense of \cite{bruit} are removed.
\end{itemize}

\subsubsection{Riachy's trick} 
It permits \cite{nicu} to avoid the calculation of the derivative $\dot{e}$ in Eqn.~\eqref{ipd}. Rewrite~\eqref{ul2} as $\ddot{y}(t) + K_D \dot{y}(t) = F(t) + K_D \dot{y}(t) + \alpha u(t)$. Set 
$$Y(t) = y(t) + K_D \int_{c}^t y(\sigma)d\sigma, \quad 0 \leqslant c < t$$
It yields $\ddot{Y}(t) = \ddot{y}(t) + K_D \dot{y}(t)$. Set $\mathcal{F}(t) = F(t) + K_D \dot{y}(t)$. Eqn. \eqref{ul2} becomes
    $\ddot{Y}(t) = \mathcal{F}(t) + \alpha u(t)$.
Eqn. \eqref{ipd} reads now
\begin{equation}\label{ipd2}
    u(t) = -\frac{\mathcal{F}_{\rm est}(t) - \ddot{y}^\star (t)  +K_P e(t) + K_D \dot{y}^\star (t)}{\alpha}
\end{equation}
where the derivative of $y$ and, therefore, $e$ disappears. The estimate $\mathcal{F}_{\rm est}$ in Eqn. \eqref{ipd2} may be computed via Formula \eqref{estim} by replacing $y$ by $Y$.



\subsection{Computer simulations}\label{simu}
\subsubsection{Presentation}
The numerical values of the parameters in Eqn. \eqref{model} are taken from 
\cite{arrais20,arrais19}: 
$C_1=135$, $C_2=0.8C_1$, $C_3=0.25C_1$, $C_4=0.25C_1$, $C_7=C_2$, $C_5=0.3C_1$, $C_6=0.1C_1$, $A=3.25$, $B=22$, $G=20$, $a=100$, $b=30$, $g=350$.
Following again \cite{arrais20,arrais19}, set
\begin{itemize}
    \item $p = 200$, $0 \leqslant t \leqslant 2{\text s}$, for a regular behavior;
    \item $p = 800$, $t > 2{\text s}$, for an abnormal behavior.
\end{itemize}
Moreover, $p$ is corrupted by an additive white Gaussian noise (mean: $0$, standard deviation: $10$).
 

The stimulation begins when the anomaly is detected. Several scenarios are now presented.

\subsubsection{Scenarios 1 and 2}
Assume that $y_1 (t)$ is available. Set $y(t) = y_1(t)$ in Eqn. \eqref{ul2}, where $\alpha = 10^4$.\footnote{This numerical value might seem huge to readers which are used to this model-free approach. It is easily explained here by looking at the second line of Eqn. \eqref{model} and computing the product $AaC_2$ in front of the sigmoid function.} Set $K_P=100$, $K_D=20$ in Eqn. \eqref{ipd}.  

\begin{figure*}[!ht]
\centering%
\subfigure[\footnotesize States ]
{\epsfig{figure=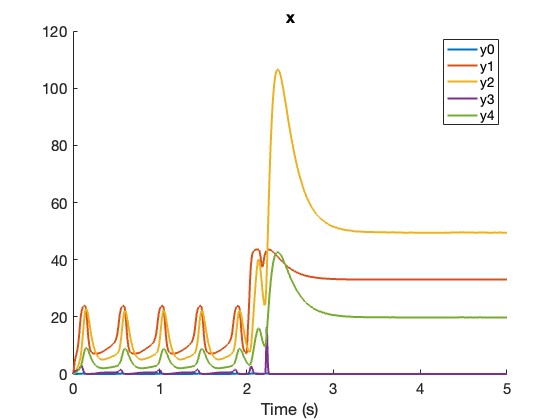,width=0.24\textwidth}}
\subfigure[\footnotesize $y_1$ (red ---) and $y^\star$ (blue - -)]
{\epsfig{figure=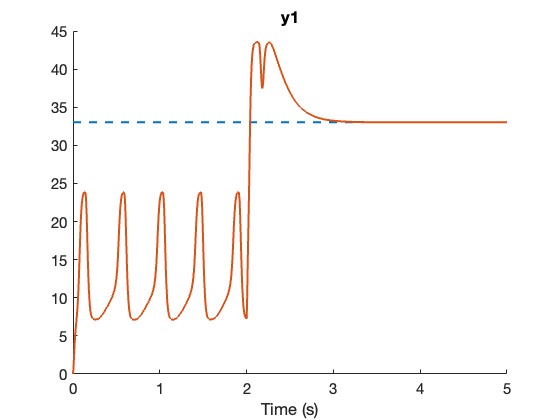,width=0.24\textwidth}}
\subfigure[\footnotesize Control input $u$ ]
{\epsfig{figure=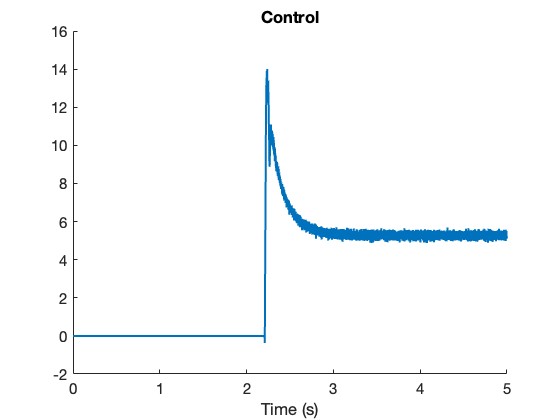,width=0.24\textwidth}}
\caption{Closed loop result: Scenario 1}\label{S1c}
\end{figure*}

\begin{figure*}[!ht]
\centering%
\subfigure[\footnotesize States ]
{\epsfig{figure=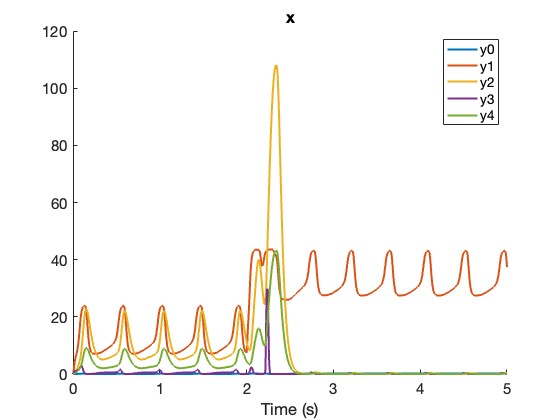,width=0.24\textwidth}}
\subfigure[\footnotesize $y_1$ (red ---) and $y^\star$ (blue - -)]
{\epsfig{figure=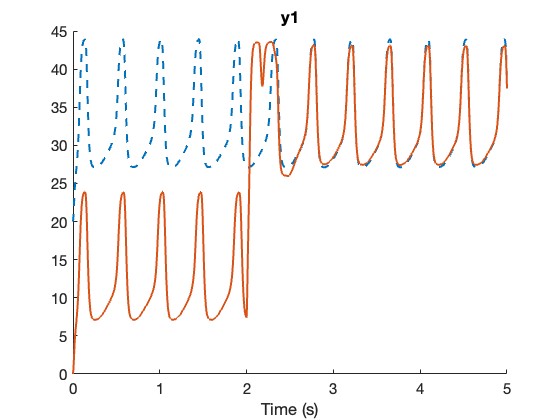,width=0.24\textwidth}}
\subfigure[\footnotesize Control input $u$ ]
{\epsfig{figure=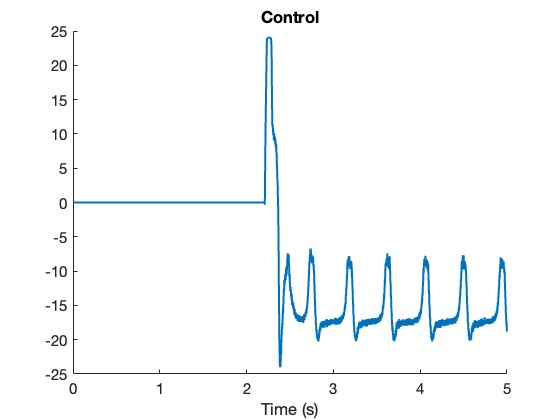,width=0.24\textwidth}}
\caption{Closed loop result: Scenario 2}\label{S2}
\end{figure*}

In Scenario 1 (Fig. \ref{S1c}) a constant reference is perfectly tracked (Fig. \ref{S1c}-(b)) with very reasonable values of the stimulation: $0 \leqslant u(t) \leqslant 15$ (Fig. \ref{S1c}-(c)). It is quite obvious to check that the very nature of the sigmoid function in Eqn.~\eqref{model} prevents to track references of arbitrary magnitude.
In Scenario 2 (Fig. \ref{S2}) the reference trajectory is reproducing a crisis-free recording (Fig. \ref{S2}-(b)). 
The tracking is again excellent.

\subsubsection{Scenario 3 -- Several virtual patients}
\begin{figure*}[!ht]
\centering%
\subfigure[\footnotesize States ]
{\epsfig{figure=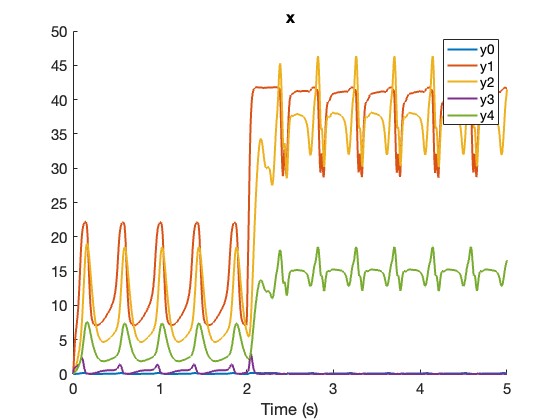,width=0.24\textwidth}}
\subfigure[\footnotesize $y_m$ (red ---) and $y^\star$ (blue - -)]
{\epsfig{figure=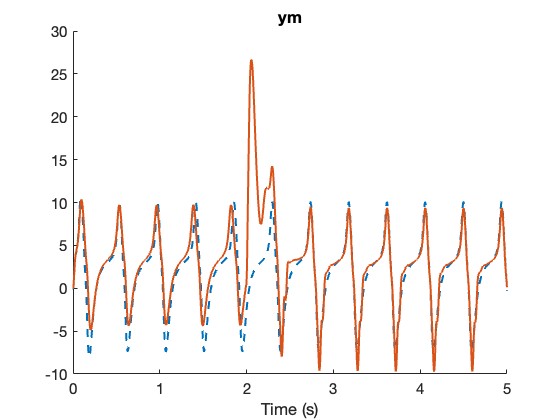,width=0.24\textwidth}}
\subfigure[\footnotesize Control input $u$ ]
{\epsfig{figure=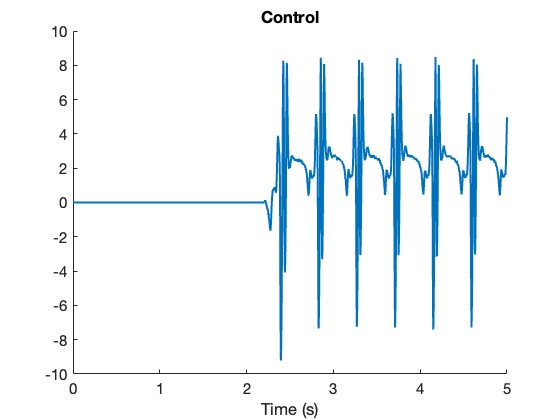,width=0.24\textwidth}}
\caption{Closed loop result: Scenario 3a}\label{S3a}
\end{figure*}

\begin{figure*}[!ht]
\centering%
\subfigure[\footnotesize States ]
{\epsfig{figure=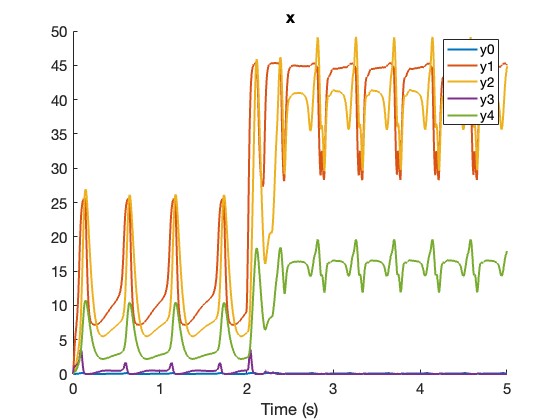,width=0.24\textwidth}}
\subfigure[\footnotesize $y_m$ (red ---) and $y^\star$ (blue - -)]
{\epsfig{figure=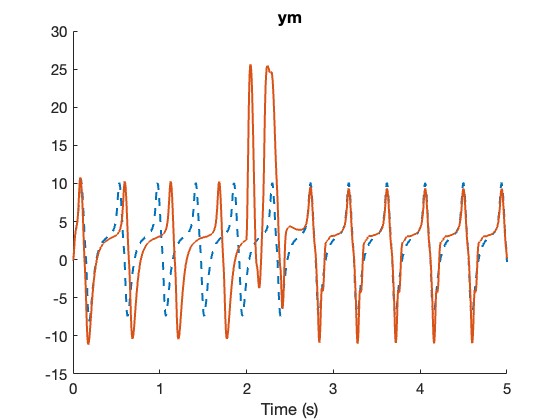,width=0.24\textwidth}}
\subfigure[\footnotesize Control input $u$ ]
{\epsfig{figure=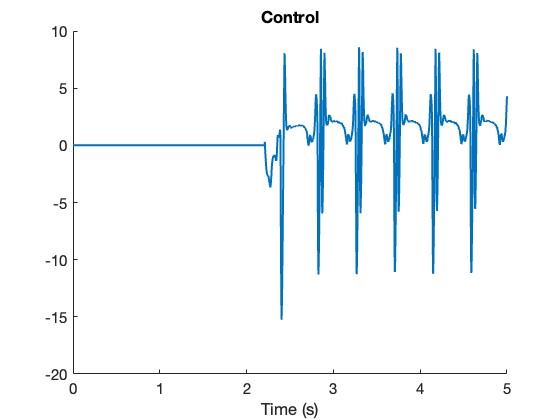,width=0.24\textwidth}}
\caption{Closed loop result: Scenario 3b}\label{S3b}
\end{figure*}

\begin{figure*}[!ht]
\centering%
\subfigure[\footnotesize States ]
{\epsfig{figure=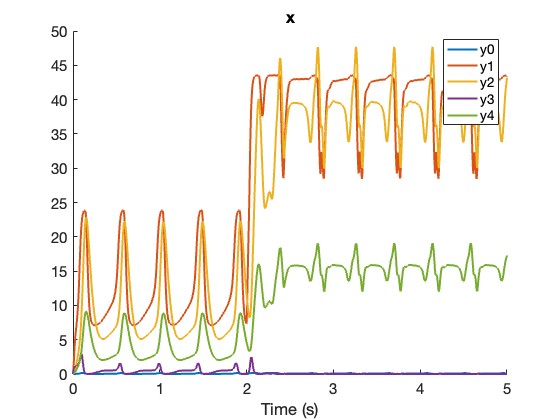,width=0.24\textwidth}}
\subfigure[\footnotesize $y_m$ (red ---) and $y^\star$ (blue - -)]
{\epsfig{figure=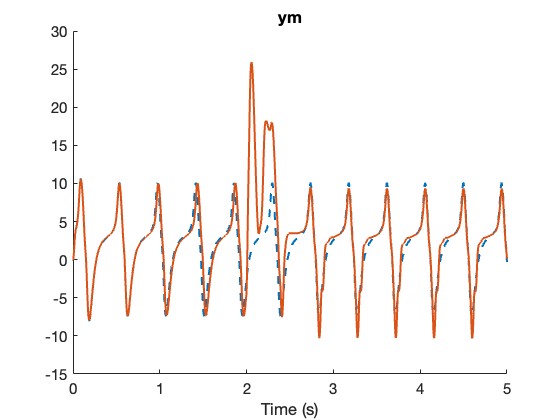,width=0.24\textwidth}}
\subfigure[\footnotesize Control input $u$ ]
{\epsfig{figure=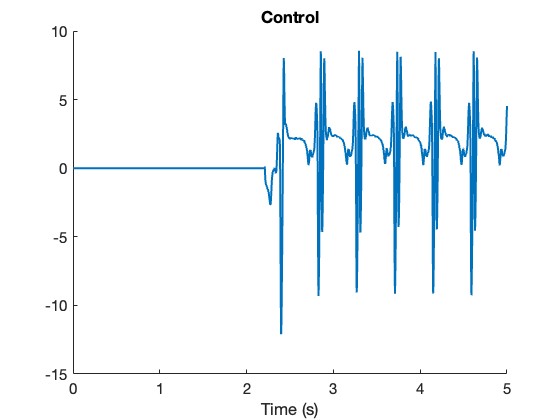,width=0.24\textwidth}}
\caption{Closed loop result: Scenario 3c}\label{S3c}
\end{figure*}

Set in Scenario 3 (Figs. \ref{S3a}, \ref{S3b}, \ref{S3c}) $y(t) = y_m (t)$,  $\alpha=-10^5$  in Eqn. \eqref{ul2}, and  $K_P=400$, $K_D=40$ in Eqn. \eqref{ipd}. The following numerical variations on $C_\iota$, $\iota = 1, \dots, 7$ correspond to different virtual patients. 
    Take $0.9 \times C_\iota$ (resp. $1.1 \times C_\iota$), $\iota = 1, \dots, 7$. Results are reported in Fig. \ref{S3a} (resp. Fig.~\ref{S3b}). See Fig. \ref{S3c} for the results with the nominal values of $C_\iota$.
    Tracking is excellent in all three cases.


\subsubsection{Scenario 4 -- Measurement noise}
\begin{figure*}[!ht]
\centering%
\subfigure[\footnotesize States ]
{\epsfig{figure=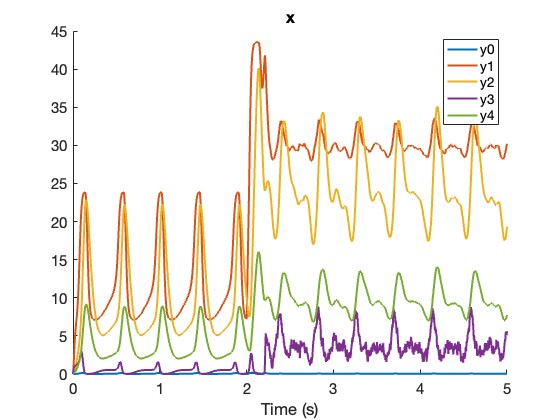,width=0.24\textwidth}}
\subfigure[\footnotesize $y_m$ (red ---) and $y^\star$ (blue - -)]
{\epsfig{figure=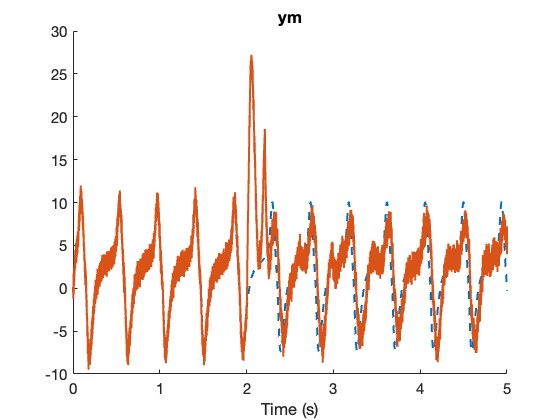,width=0.24\textwidth}}
\subfigure[\footnotesize Control input $u$ ]
{\epsfig{figure=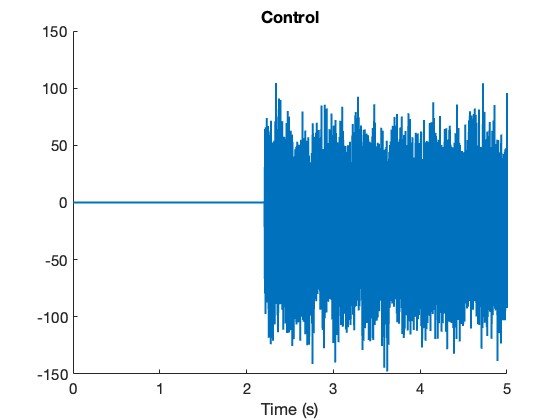,width=0.24\textwidth}}
\caption{Closed loop result: Scenario 4}\label{S4}
\end{figure*}

Again $y(t) = y_m (t)$,  $\alpha=-10^5$  in Eqn. \eqref{ul2}, and  $K_P=400$, $K_D=40$ in Eqn. \eqref{ipd2}. In Scenario 4 (Fig. \ref{S4}) an additive white Gaussian noise has been added as a measurement noise (mean: 0, standard deviation: 0.5). 
Performances deteriorate only slightly with a chattering control variable. 

\section{Conclusion}\label{conclu}
Our communication suggests that some recent control techniques, which were already most useful in industry, might also be helpful for implementing closed-loop neurostimulations for curing epileptiform seizures. Even if our results seem, to the best of our knowledge, to surpass the existing literature, there is still a lot of work to be done to go beyond virtual patients. Future publications will soon examine other topics in neuroscience, like Parkinson's disease and brain plasticity.


\bibliographystyle{abbrv}

\end{document}